\documentclass[conference]{IEEEtran}
\IEEEoverridecommandlockouts

\usepackage{cite}
\usepackage{amsmath,amssymb,amsfonts}
\usepackage{algorithmic}
\usepackage{graphicx}
\usepackage{textcomp}
\usepackage{xcolor}
\usepackage{amsmath,graphicx,booktabs,multirow}

\def\BibTeX{{\rm B\kern-.05em{\sc i\kern-.025em b}\kern-.08em
    T\kern-.1667em\lower.7ex\hbox{E}\kern-.125emX}}
\begin{document}

\title{Adopting Whisper for Confidence Estimation}

\author{\IEEEauthorblockN{Vaibhav Aggarwal$^\dagger$ \thanks{$^\dagger$Equal Contribution.}, Shabari S Nair$^\dagger$, Yash Verma$^\dagger$, Yash Jogi}
\IEEEauthorblockA{\textit{Sprinklr} \\
Gurugram, India \\
\texttt{\{vaibhav.aggarwal, shabari.nair, yash.verma, yash.jogi\}@sprinklr.com}
}}
\maketitle

\begin{abstract}

Recent research on word-level confidence estimation for speech recognition systems has primarily focused on lightweight models known as Confidence Estimation Modules (CEMs), which rely on hand-engineered features derived from Automatic Speech Recognition (ASR) outputs. In contrast, we propose a novel end-to-end approach that leverages the ASR model itself (Whisper) to generate word-level confidence scores. Specifically, we introduce a method in which the Whisper model is fine-tuned to produce scalar confidence scores given an audio input and its corresponding hypothesis transcript. Our experiments demonstrate that the fine-tuned Whisper-tiny model, comparable in size to a strong CEM baseline, achieves similar performance on the in-domain dataset and surpasses the CEM baseline on eight out-of-domain datasets, whereas the fine-tuned Whisper-large model consistently outperforms the CEM baseline by a substantial margin across all datasets.

\end{abstract}

\begin{IEEEkeywords}
confidence, Whisper, automatic speech recognition, calibration
\end{IEEEkeywords}

\section{Introduction}

Confidence estimation plays a crucial role in assessing the reliability of transcripts  generated by end-to-end (E2E) ASR models \cite{jiang2005confidence, qiu2021learning}. Word-level confidence estimators assign a probability score to each word of the hypothesis transcript, indicating its likelihood of being accurate \cite{qiu2021learning, qiu2021multi}. These confidence scores are valuable for various applications, for instance, selecting utterances for training in semi-supervised learning \cite{tur2005combining, huang13b_interspeech}, and for flagging  and reviewing potential word-errors in the transcripts \cite{qiu2021learning, naowaratword}.

A simple method to calculate confidence for E2E neural-network models is by using the model output probabilities themselves \cite{hendrycks2017a}. For example, for E2E ASR models, word-level confidence can be estimated by simply aggregating the model output probabilities of each token of the word \cite{qiu2021learning, naowaratword}. This method is simple, fast, and easy to implement. However, this approach often suffers from overconfidence issue, wherein the model tends to output high probabilities for words that are actually incorrect \cite{laptev2023fast}, thereby reducing the reliability  of this method for confidence estimation. 

To address this issue, previous work \cite{qiu2021learning, li2021confidence} has introduced additional features derived from model outputs, which are processed by a neural-network known as the Confidence Estimation Module (CEM) to produce word-level confidence estimates. For instance, in \cite{qiu2021learning}, an attention-based CEM, utilizing features such as top-$K$ probabilities and hidden states from the last decoder layer, outperforms the method that relies on model probabilities alone. Moreover, recent research has demonstrated several advancements in enhancing the capabilities of CEM. For example, to enhance CEM's performance on out-of-domain (OOD) data, the work of \cite{li2022improving} demonstrated that further training on unlabeled OOD audio data, combined with features from a language model trained on OOD textual data, significantly improves performance. Additionally, \cite{qiu2021multi} extended CEM to output number of deletions as well as utterance-level confidence, while \cite{naowaratword} built a similar CEM for Connectionist Temporal Classification (CTC) \cite{graves2006connectionist} models such as wav2vec2.0 \cite{baevski2020wav2vec}. Overall, these approaches enhance confidence estimation by incorporating diverse hand-engineered features into light-weight models.

In \cite{lin2022teaching}, it is demonstrated that GPT-3 (Generative Pre-trained Transformers 3) \cite{brown2020language}, a large language model, can be fine-tuned to express confidence for its own answers in natural language. Specifically, GPT-3, originally pre-trained for next-token prediction, was fine-tuned to not only generate answers for input prompts but also to provide confidence scores in textual form. Inspired by this aspect of the work, wherein the model itself is fine-tuned for confidence estimation in an end-to-end fashion, we ask an intriguing question: Can an ASR model be fine-tuned to predict word-level confidence for its own hypothesis transcripts? 

To test this hypothesis, we use a widely-used foundation model for speech--Whisper\cite{radford2023robust}. We propose a simple yet novel approach to utilize Whisper as a word-level confidence estimator. Particularly, we showcase a methodology wherein we create a replica of pre-trained Whisper and fine-tune its decoder to generate word-level confidence scores for an input audio and its hypothesis transcript (generated by the Whisper-large model). We name the fine-tuned model \textit{C-Whisper}.

Our experiments showcase that C-Whisper initialized from the Whisper-tiny model performs comparably with CEM on the in-domain dataset but outperforms on the OOD datasets, whereas C-Whisper initialized from the Whisper-large model surpasses CEM by a large margin on all the datasets. 
Additionally, we show that C-Whisper achieves superior performance even on an out-of-the-box ASR service, indicating its ability to be adopted for other ASR systems.

\section{Background}

\textbf{CEM: }For the baseline in our study, we use the CEM implementation as described in \cite{qiu2021learning}, since it has demonstrated superior performance compared to the basic model probabilities method and has influenced subsequent research \cite{qiu2021multi, li2022improving, naowaratword}. 

The CEM approach in \cite{qiu2021learning} was implemented for the Recurrent Neural Network Transducer (RNN-T) ASR architecture \cite{he2019streaming, sainath2020streaming}. However, it can be applied to any sub-word E2E ASR system, such as Whisper.

The main inputs to the CEM are derived from the ASR model outputs. Specifically, for each token in the hypothesis transcript, a feature vector $F$ is constructed. This vector $F$ consists of the token's log probability, the top-$K$ log probabilities from the output probability distribution, the token's embedding (with added positional embedding), and the last hidden states of the decoder. Additional inputs include the acoustic features ($E$) extracted from the ASR model encoder and the linguistic features extracted from the top-$N$ beam-search hypotheses ($B$).

During a single forward pass of the CEM, $F$ is processed through a self-attention layer to produce $F'$. This is then cross-attended with both the encoder features $E$ and the beam-search hypotheses $B$. The outputs of these two cross-attention layers are concatenated and passed through a linear layer to generate confidence scores for each token. Finally, word-level confidence scores are derived by taking the confidence score of the last token of each word. For more details on CEM, refer to \cite{qiu2021learning}.

\textbf{Whisper: }Whisper is a speech foundation model pre-trained on $680,000$ hours of data for various speech-related tasks such as speech transcription, speech translation, etc. The model follows a Transformer \cite{vaswani2017attention} based Encoder-Decoder architecture. 

\begin{figure}[!t]
\centerline{{\includegraphics{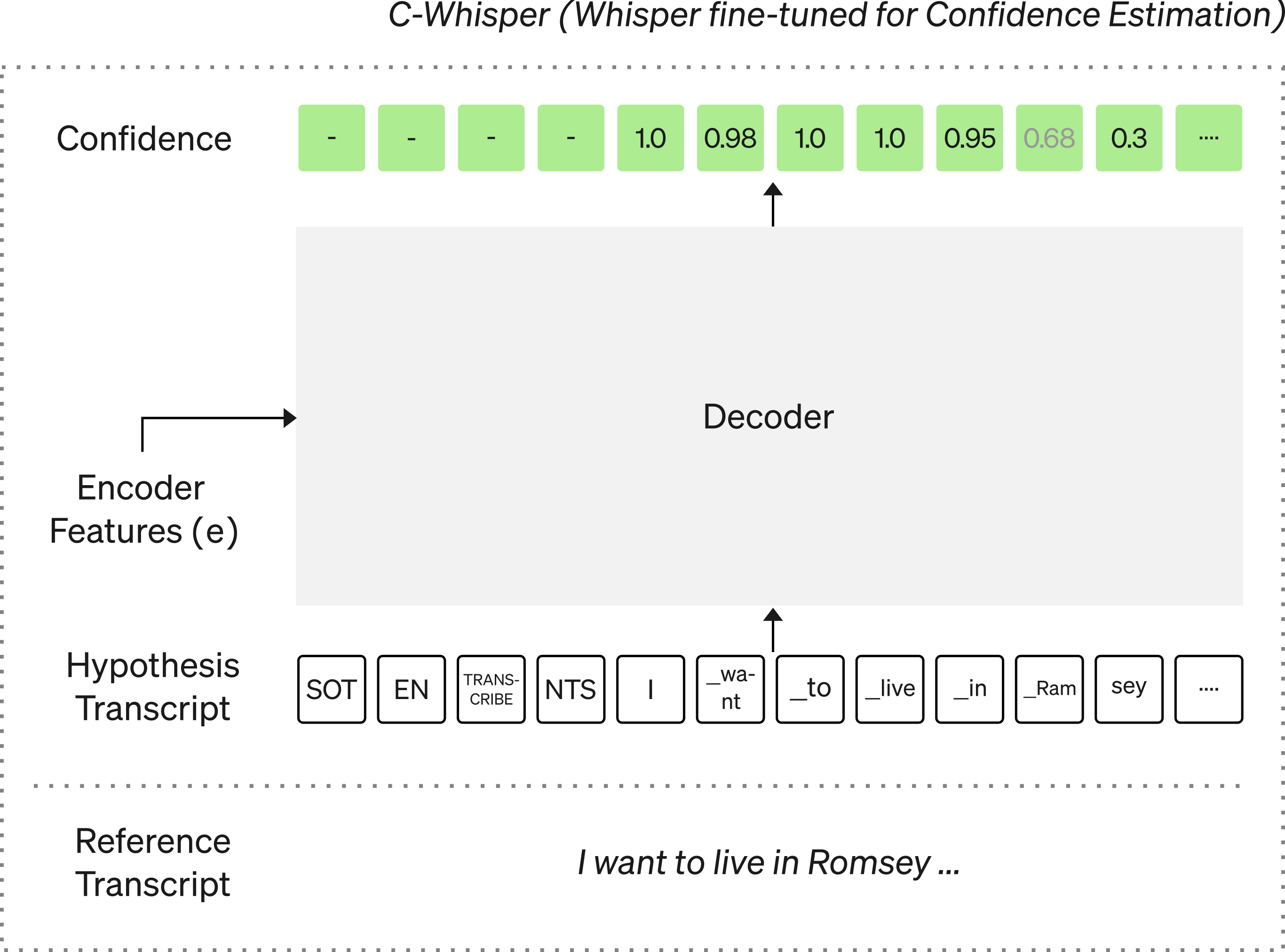}}}
\hfill
\caption{C-Whisper: The model's decoder takes both the hypothesis transcript and encoder features as inputs and produces token-level confidence scores. To represent the confidence for an entire word, we use the confidence score of the last token of each word. Consequently, the confidence scores for tokens that do not occur at the end of a word are displayed in gray color.}
\label{fig}
\end{figure}

Given an audio input $A$, the Whisper Encoder processes $A$ to generate its corresponding encoder features $e$. The Whisper decoder $D$ then utilizes this $e$ and the previously generated tokens $t_{<i}$ to output the hidden state $h_i$, where $i$ is the current decoding step. This can be represented by the following equation:

\begin{equation}
    h_i = D(e, t_{<i})
\label{fig:eq1}
\end{equation}

The hidden state $h_i$ is passed through a linear layer to produce the logits distribution $l_i$ over the sub-word vocabulary of the ASR model. This distribution $l_i$ is then used to predict the next token. The relationship is defined by the following equation:

\begin{equation} l_i = W_{w}^{T}h_i + B_{w} 
\label{fig:eq2}
\end{equation}
\noindent
where $W_w$ and $B_w$ denote the weights and biases of the linear layer, respectively. For more information related to Whisper, refer to \cite{radford2023robust}.

\section{Methodology}

In this section, we describe the methodology to fine-tune the Whisper model to output word-level confidence scores.

As described previously, the Whisper decoder essentially merges audio input (encoder features) with text input (previously decoded tokens) to output text (hypothesis transcript). However, for word-level confidence estimation, the task is to output confidence scores (scalar values), given the audio input and its hypothesis transcript (text). Therefore, to adapt Whisper for confidence estimation via transfer learning, the decoder needs to be modified to output scalar values instead of next-token probability distribution. To implement this change, we remove the last linear layer of the decoder and replace it with a newly initialized layer that maps $h_i$ to a single scalar value, depicted by the following equation:

\begin{equation}
    c(t_i) = \sigma(W_{c}^{T}{h_i} + B_{c})
\label{fig:eq3}
\end{equation}
\noindent 
where $t_i$ is the $i^{th}$ token of the hypothesis transcript, $c(t_i)$ represents the confidence for the token $t_i$, $W_c$ and $B_c$ denote the weights and biases of the newly initialized linear layer respectively, $\sigma$ is the sigmoid function and $h_i$ is as defined as (\ref{fig:eq2}). In this slightly-modified architecture, the hypothesis transcript $t$ is fed back to the decoder to output token level confidence values. 
As a supplementary point, it is worth noting that this approach of fine-tuning a model originally pre-trained for next-token prediction to perform scalar prediction has been employed in other contexts as well. Specifically, it has been used to adapt large language models (LLMs) to function as reward models \cite{ouyang2022training}.

Finally, to generate word-level confidence scores from these token-level scores, we adopt the approach from \cite{qiu2021learning}, wherein the last token's confidence is considered to be the confidence for a given word. This method of aggregation proved to be the most optimal in our experiments, compared to other methods such as minimum, product and mean of word-piece confidences.

One significant implication of this architectural change is that it makes the decoder non-auto-regressive, since the tokens are processed in parallel instead of sequentially. A concern that can arise here is that, while training, the Whisper decoder uses a causal attention mask to restrict a token from attending to future tokens, preserving its auto-regressive nature. However, for word-level confidence estimation, it could be advantageous for each token to attend to all the tokens in the hypothesis transcript to allow for more comprehensive learning. To investigate this, we performed two experiments: one retaining the causal attention mask and one without it. Surprisingly, the results indicated that retaining the causal attention mask led to better performance. Further details on this are provided in Section V.

We fine-tuned the modified model using pairs of raw audio and their corresponding hypothesis transcripts as input, aiming to predict token-level confidence scores as outputs. The model was fine-tuned using Binary Cross-Entropy (BCE) Loss on the predicted word-level confidence scores, using ground truth binary labels created by aligning the reference and hypothesis transcripts (in a manner similar to that outlined in \cite{qiu2021learning}). The encoder was kept frozen during the fine-tuning process. Figure \ref{fig} depicts the workings of our fine-tuned model \textit{C-Whisper}.

\textbf{Comparison with CEM: } When comparing C-Whisper with CEM, several key differences emerge. The first  key difference is in the type of inputs: CEM expects various hand-picked features derived from model outputs, while C-Whisper expects just two inputs—audio and its hypothesis transcript. This makes C-Whisper an end-to-end confidence estimator, and hence, making it model-independent. 

The second key difference is in how the model parameters are initialized. C-Whisper is initialized with the parameters of Whisper (except the newly added linear layer)--a model pre-trained on astronomical $680K$ hours of data sourced from the internet. In contrast, CEM is initialized from scratch without any pre-training. This difference can affect each model’s ability to generalize to out-of-distribution (OOD) datasets.

\section{Experimental Setup}

\textbf{Train and Test sets:} For training all the models, we used the Common Voice $18$ English \cite{cv} train set, which comprises of approximately $1,700$ hours of audio-text pairs.

In addition to evaluating the models on the Common Voice test set, we assessed their performance on several other datasets to benchmark the generalization capabilities in OOD conditions. These datasets include the test sets of LibriSpeech-Clean (LS-Clean) \cite{libri}, LibriSpeech-Other (LS-Other), Voxpopuli English \cite{vox}, Chime6 \cite{chime}, Fleurs English \cite{fleurs}, and Multilingual LibriSpeech dataset \cite{pratap20_interspeech}, specifically for three languages: French (MLS-Fr), Italian (MLS-It), and Portuguese (MLS-Pt). Note that none of the datasets mentioned above were used during pre-training of the Whisper model.

\textbf{Model Details:} In all our experiments, we chose the Whisper-large model as the base ASR to generate hypothesis transcripts. The Word Error Rate (WER) of the hypotheses generated by Whisper is as follows for each dataset (in order as presented in Table \ref{tab:performance-metrics}): $10.1\%$, $2.6\%$, $5.3\%$, $7.1\%$, $25.5\%$, $6.3\%$, $8.6\%$, $17.7\%$, and $7.94\%$. 

For confidence estimation using raw model probabilities, we calculate word-level confidence by taking the minimum of the softmax probabilities assigned to the constituent tokens of a word. This approach outperformed in comparison to other common methods such as mean, sum, product, and maximum. We refer to this method as ``Softmax'' throughout the rest of the paper.

For the CEM implementation, we adopted the architecture and hyper-parameters as mentioned in \cite{qiu2021learning}. As the original closed-source BiLSTM encoder \cite{delib} in the deliberation model was unavailable, we replaced it with the open-source \texttt{all-MiniLM-L12-v2} \cite{sentence_emb} (fine-tuned on top of widely-used mini-language-model  \cite{wang2020minilm}), an efficient $33$M parameter model that generates a $384$-dimensional sentence embedding for each beam hypothesis.

In the case of C-Whisper we trained two versions: C-Whisper-tiny and C-Whisper-large, initialized from Whisper-tiny and Whisper-large, respectively. Despite having four decoder layers, C-Whisper-tiny ($39M$) is smaller than the CEM model ($96M$) having only one transformer layer, even without considering the sentence encoder. Both the models were trained with a learning rate of $5 \times 10^{-6}$ for 1 epoch using the Adam optimizer and a linear learning rate decay, with $10\%$ dropout rate. 

\textbf{Metrics: } We evaluate word-level confidence models using several popular metrics \cite{metrics, li2021confidence, qiu2021learning, li2022improving}. NCE (Normalized Cross Entropy) \cite{Siu1997ImprovedEE} indicates how close the confidence values for incorrect words are to $0$ and for correct words to $1$, and it ranges from $-\infty$ to $1$. Before computing NCE, confidence values are calibrated using histogram binning method \cite{calibration}. AUC-ROC (Area under the Receiver Operating Characteristic curve) measures the correlation between confidence predictions and word correctness, and ranges from $0$ to $1$. Due to AUC-ROC's limitations with imbalanced class distribution \cite{rocbad}, we also benchmark the models on AUC-PR (Area under the Precision-Recall curve). Given Whisper’s high accuracy and minority of errors, we report on two variations of AUC-PR: AUC-PR$_{POS}$ (traditional AUC-PR) and AUC-PR$_{NEG}$ (AUC-PR with errors as positives). Higher values indicate better performance for all the metrics.

\section{Results and Discussion}

\begin{table*}[t]

\caption{Confidence metrics for Softmax, CEM, C-Whisper-tiny and C-Whisper-large across different datasets}
\label{tab:performance-metrics}
\resizebox{\textwidth}{!}{
\begin{tabular}{@{}l|l|cccccc|ccc@{}}
\toprule
Metric & Model & CV & LS-Clean & LS-Other & VoxPopuli & Chime6 & Fleurs & MLS-Fr & MLS-It & MLS-Pt \\
\midrule
\multirow{5}{*}{NCE ($\uparrow$)} & Softmax & 0.313 & 0.163 & 0.183 & 0.120 & 0.103 & 0.159 & 0.106 & 0.099 & 0.057 \\
 & CEM & 0.389 & 0.253 & 0.302 & 0.126 & 0.143 & 0.179 & 0.113 & 0.077 & 0.095 \\
 & C-Whisper-tiny & 0.388 & 0.431 & 0.338 & 0.167 & 0.332 & 0.192 & 0.223 & 0.056 & 0.164 \\

 & C-Whisper-large & \textbf{0.541} & \textbf{0.502} & \textbf{0.455} & \textbf{0.226} & \textbf{0.411} & \textbf{0.257} & \textbf{0.380} & \textbf{0.281} & \textbf{0.257} \\
\midrule
\multirow{5}{*}{AUC-ROC ($\uparrow$)} & Softmax & 0.862 & 0.783 & 0.809 & 0.736 & 0.694 & 0.764 & 0.711 & 0.728 & 0.794 \\
 & CEM & 0.884 & 0.799 & 0.849 & 0.706 & 0.732 & 0.735 & 0.709 & 0.705 & 0.732 \\
 & C-Whisper-tiny & 0.897 & 0.887 & 0.870 & 0.733 & 0.865 & 0.785 & 0.810 & 0.682 & 0.774 \\

 & C-Whisper-large & \textbf{0.944} & \textbf{0.922} & \textbf{0.920} & \textbf{0.773} & \textbf{0.898} & \textbf{0.828} & \textbf{0.896} & \textbf{0.851} & \textbf{0.863} \\
\midrule
\multirow{5}{*}{AUC-PR$_{POS}$ ($\uparrow$)} & Softmax & 0.980 & 0.992 & 0.986 & 0.980 & 0.856 & 0.981 & 0.934 & 0.925 & 0.976 \\
 & CEM & 0.984 & 0.993 & 0.989 & 0.977 & 0.850 & 0.981 & 0.939 & 0.915 & 0.966 \\
 & C-Whisper-tiny & 0.985 & 0.996 & 0.990 & 0.978 & 0.937 & 0.982 & 0.951 & 0.900 & 0.968 \\

 & C-Whisper-large & \textbf{0.992} & \textbf{0.997} & \textbf{0.994} & \textbf{0.983} & \textbf{0.953} & \textbf{0.987} & \textbf{0.976} & \textbf{0.954} & \textbf{0.983} \\
\midrule
\multirow{5}{*}{AUC-PR$_{NEG}$ ($\uparrow$)} & Softmax & 0.558 & 0.204 & 0.319 & 0.220 & 0.533 & 0.297 & 0.380 & 0.395 & 0.351 \\
 & CEM & 0.643 & 0.357 & 0.464 & 0.279 & 0.602 & 0.370 & 0.397 & 0.372 & 0.252 \\
 & C-Whisper-tiny & 0.628 & 0.580 & 0.519 & 0.345 & 0.749 & 0.364 & 0.530 & 0.337 & 0.394 \\

 & C-Whisper-large & \textbf{0.769} & \textbf{0.643} & \textbf{0.616} & \textbf{0.423} & \textbf{0.811} & \textbf{0.404} & \textbf{0.613} & \textbf{0.649} & \textbf{0.437} \\
\bottomrule
\end{tabular}
}
\end{table*}
\begin{table}[!t]
\caption{Average confidence metrics across all English datasets with and without causal attention mask}
\centering
\small
\resizebox{0.95\columnwidth}{!}{
\begin{tabular}{@{}l|*{4}{c}@{}}
\toprule
Metric & NCE & AUC-ROC & AUC-PR$_{POS}$ & AUC-PR$_{NEG}$ \\ 
\midrule
C-Whisper-large & \textbf{0.399} & \textbf{0.881} & \textbf{0.984} & \textbf{0.611} \\ 
+ Non-Causal Attention & 0.321 & 0.854 & 0.979 & 0.529 \\ 

C-Whisper-tiny & \textbf{0.308} & \textbf{0.840} & \textbf{0.978} & \textbf{0.531} \\ 
+ Non-Causal Attention & 0.222 & 0.789 & 0.971 & 0.433 \\ 
\bottomrule
\end{tabular}%
}

\label{tab:model-comparison-transposed}
\end{table}

\begin{table}[htbp]
\caption{Confidence metrics for Company-x asr on Librispeech dataset}
\centering
\small
\resizebox{0.95\columnwidth}{!}{%
\begin{tabular}{@{}l|*{4}{c}@{}}
\toprule
Model & NCE & AUC-ROC & AUC-PR$_{POS}$ & AUC-PR$_{NEG}$ \\
\midrule
ASR-CEM & 0.296 & 0.862 & 0.979 & 0.510 \\
C-Whisper-tiny & 0.219 & 0.821 & 0.972 & 0.429 \\
C-Whisper-large & \textbf{0.526} & \textbf{0.938} & \textbf{0.991} & \textbf{0.747} \\
\bottomrule
\end{tabular}%
}
\label{tab:librispeech-other-comparison}
\end{table}

Table \ref{tab:performance-metrics} presents the performance of various models across all the test datasets. The CEM model consistently outperforms the Softmax method for all the datasets, highlighting the performance gain obtained by utilizing ASR-output related features in addition to output probabilities for confidence estimation. Although CEM, with its 96 million parameters, achieves strong results, C-Whisper-tiny delivers comparable performance on the in-domain dataset Common Voice while outperforming it on nearly all other datasets, despite having only 39 million parameters. This improvement over CEM and Softmax on the OOD datasets can be attributed to Whisper's extensive pre-training on a large-scale web dataset, which enables the fine-tuned model to generalize effectively across various OOD datasets. In contrast, CEM may have learned spurious patterns during training specific to the Common Voice dataset, resulting in reduced performance on the OOD datasets.

Across all the datasets, C-Whisper-large achieves the best performance on all the evaluation metrics compared to the other models. However, this increase in performance comes at the cost of higher latency and a significantly larger model size. Therefore, although C-Whisper-tiny offers good performance with a fast inference time, tasks that require higher accuracy, such as active learning \cite{riccardi2005active}---where inference time is less critical---may benefit from the use of C-Whisper-large.

To evaluate the impact of retaining the causal attention mask in the decoder, we conducted experiments with C-Whisper-large and C-Whisper-tiny. The average results of these experiments are presented in Table \ref{tab:model-comparison-transposed}. In both models,
replacing the causal attention mask with non-causal attention, surprisingly, led to a drop in performance. However, our findings reveal that the performance gap between the causal and non-causal models diminishes as training progresses. Specifically, in the early stages of training, the causal model significantly outperforms the non-causal model. This is likely due to the models being initialized from the pre-trained Whisper-large checkpoint, which was originally trained with a causal attention mask. As training continues, the performance gap narrows, suggesting that the non-causal model adapts more effectively over time. In future work, we plan to investigate these dynamics further to better understand the underlying reasons for this observed behavior.

Moreover, to evaluate C-Whisper's performance for another ASR system, we experimented on an out-of-the-box ASR model of the Company-X Transcribe service. Table \ref{tab:librispeech-other-comparison} shows the performance of the confidence estimates of the Company-X ASR (labeled as `ASR-CEM'), C-Whisper-tiny, and C-Whisper-large on the Libri-Other dataset. It is clear that though C-Whisper-tiny falls short of ASR-CEM, C-Whisper-large surpasses it by a large margin on all the metrics. This proves that unlike most other CEMs in the past literature, which are constrained to work only for a single ASR due to their architectural constraint, C-Whisper is able to perform well as a confidence estimator, not only for Whisper, but also for another ASR model in a zero-shot manner. Thus, it can be easily adopted for other ASR models, possibly without any modification or fine-tuning.

\section{Conclusion}

In this paper, we introduced a simple yet novel approach for word confidence estimation by adopting the ASR model itself for the task. Our method shows substantial improvements in confidence estimation compared to traditional techniques such as CEM, especially when applied to out-of-domain datasets. In the future work, we plan to extend our approach to additional ASR architectures and aim to develop a unified ASR model that not only generates transcripts but also provides corresponding confidence scores for each word in natural language.

\section*{Acknowledgment}

We thank Aayush Kubba and Yoginkumar Patel for their valuable feedback and guidance throughout this research.

\bibliographystyle{IEEEtran}
\bibliography{strings}

\end{document}